%% file: main.tex
\documentclass[%
reprint,
superscriptaddress,
amsmath,amssymb,
prb,
]{revtex4-2}

\usepackage{graphicx}% Include figure files
\usepackage{dcolumn}% Align table columns on decimal point

\usepackage{lipsum}

\usepackage{hyperref}
\usepackage{xcolor}
\hypersetup{
    colorlinks,
    linkcolor={blue},
    citecolor={blue},
    urlcolor={blue}
}

\usepackage{placeins}

\usepackage{float}

\usepackage{tikz}
\usetikzlibrary{patterns}

\newcommand*{\ditto}{---\texttt{"}---}

\newcommand*{\mt}{m^t}
\newcommand*{\mz}{m^z}
\newcommand*{\mdos}{m^\text{dos}}

\usepackage[normalem]{ulem}

\usepackage{capt-of}
\begin{document}

% \preprint{APS/123-QED}
\title{Holes in silicon are heavier than expected: transport properties of extremely high mobility electrons and holes in silicon MOSFETs}

% AUTHOR LIST (FULL NAMES)
% \author{J. P. Wendoloski}
% \author{J. Hillier}
% \author{M. Rendell}
% \author{Y. Ashlea-Alava}
% \author{S. Liles}
% \affiliation{School of Physics, University of New South Wales, New South Wales 2052, Australia}

% \author{Bart Raes}
% \author{R. Li }
% \altaffiliation{now at Applied Materials, Inc.}
% \author{Stefan Kubicek}
% \author{Clement Godfrin}
% \author{Julien Jussot}
% \author{Sofie Beyne}
% \author{Danny Wan}
% \affiliation{IMEC, Remisebosweg 1, B-3001 Leuven, Belgium}

% \author{Md. Mamunur Rahman}
% \affiliation{School of Electrical Engineering and Telecommunications, University of New South Wales,  New South Wales 2052, Australia}
% \author{Steve Yianni}
% \author{Kok Wai Chan}
% \author{Fay E. Hudson}
% \author{Wee Han Lim}
% \affiliation{School of Electrical Engineering and Telecommunications, University of New South Wales,  New South Wales 2052, Australia}
% \affiliation{Diraq, Sydney, NSW, Australia}

% \author{Kristiaan De Greve}
% \affiliation{IMEC, Remisebosweg 1, B-3001 Leuven, Belgium}
% \affiliation{Department of Electrical Engineering (ESAT), KU Leuven, Leuven, Belgium}
% \author{Andrew S. Dzurak}
% \affiliation{School of Electrical Engineering and Telecommunications, University of New South Wales,  New South Wales 2052, Australia}
% \affiliation{Diraq, Sydney, NSW, Australia}
% \author{A. R. Hamilton}
% \email{Alex.Hamilton@unsw.edu.au}
% \affiliation{School of Physics, University of New South Wales, New South Wales 2052, Australia}

% AUTHOR LIST (ABBREVIATED)
\author{J. P. Wendoloski}
\author{J. Hillier}
\author{S. D. Liles}
\author{M. Rendell}
\author{Y. Ashlea-Alava}
\affiliation{School of Physics, University of New South Wales, New South Wales 2052, Australia}

\author{B. Raes}
\author{R. Li }
\altaffiliation{now at Applied Materials, Inc.}
\author{S. Kubicek}
\author{C. Godfrin}
\author{J. Jussot}
\author{S. Beyne}
\author{D. Wan}
\affiliation{IMEC, Remisebosweg 1, B-3001 Leuven, Belgium}

\author{Md. M. Rahman}
\affiliation{School of Electrical Engineering and Telecommunications, University of New South Wales,  New South Wales 2052, Australia}
\author{S. Yianni}
\affiliation{School of Electrical Engineering and Telecommunications, University of New South Wales,  New South Wales 2052, Australia}
\affiliation{Diraq, Sydney, NSW, Australia}
\affiliation{Université Grenoble Alpes, CEA, Grenoble INP, IRIG, PHELIQS, 38000 Grenoble, France}
% \altaffiliation{Université Grenoble Alpes, CEA, Grenoble INP, IRIG, PHELIQS, 38000 Grenoble, France.}
\author{K. W. Chan}
\author{F. E. Hudson}
\author{W. H. Lim}
\affiliation{School of Electrical Engineering and Telecommunications, University of New South Wales,  New South Wales 2052, Australia}
\affiliation{Diraq, Sydney, NSW, Australia}

\author{K. De Greve}
\affiliation{IMEC, Remisebosweg 1, B-3001 Leuven, Belgium}
\affiliation{Department of Electrical Engineering (ESAT), KU Leuven, Leuven, Belgium}
\author{A. S. Dzurak}
\affiliation{School of Electrical Engineering and Telecommunications, University of New South Wales,  New South Wales 2052, Australia}
\affiliation{Diraq, Sydney, NSW, Australia}
\author{A. R. Hamilton}
\email{Alex.Hamilton@unsw.edu.au}
\affiliation{School of Physics, University of New South Wales, New South Wales 2052, Australia}

% Useful papers for author information
% https://arxiv.org/pdf/2410.15590
% https://arxiv.org/pdf/2501.17814

\date{\today}

\begin{abstract}

The quality of the silicon-oxide interface plays a crucial role in fabricating reproducible silicon spin qubits. In this work we characterize interface quality by performing mobility measurements on silicon Hall bars. We find a peak electron mobility of nearly $40,000\,\text{cm}^2/\text{Vs}$ in a device with a $21\,\text{nm}$ oxide layer, and a peak hole mobility of about $2,000\,\text{cm}^2/\text{Vs}$ in a device with $8\,\text{nm}$ oxide, the latter being the highest recorded mobility for a p-type silicon MOSFET. Despite the high device quality, we note an order-of-magnitude difference in mobility between electrons and holes. By studying additional n-type and p-type devices with identical oxides, and fitting to transport theory, we show that this mobility discrepancy is due to valence band nonparabolicity. The nonparabolicity endows holes with a density-dependent transverse effective mass ranging from $0.6m_0$ to $0.7m_0$, significantly larger than the usually quoted bend-edge mass of $0.22m_0$. Finally, we perform magnetotransport measurements to extract momentum and quantum scattering lifetimes.

\end{abstract}

\maketitle

\section{Introduction}

Silicon quantum dots offer a promising path towards scalable quantum computing. This is due to their compatibility with existing semiconductor fabrication techniques, as such spin qubits are created using gate-defined potential wells in metal-oxide-semiconductor (MOS) interfaces. The spins of both electrons and holes have been used within this approach. Electron spins have received significant attention due to their long coherence times in silicon \cite{veldhorst2015two,dzurak2021scaling,huang2024high,maurand2016cmos,veldhorst2014addressable,spruijtenburg2018fabrication}. Likewise, holes show promise due to their intrinsic spin-orbit coupling, enabling fast all-electrical control of the spin state \cite{holes2007edsr,holes2018matrix,scott2018six,froning2021ultrafast,liles2021electrical,camenzind2022hole,jin2023combining,holes2023coupling,liles2023singlet,holes2024scott}. In both cases, this approach requires the fabrication of millions of these qubits, which relies on the ability to create very low disorder MOS interfaces.\\

Transport measurements performed on Hall bars have been used to characterise the disorder of n-type MOS interfaces at low temperatures \cite{characterisation_high_quality, manufacturing, industrial_quantum_transport}. The charge carrier mobility is a key metric used to assess interface quality, and theoretical models can be used to infer interface characteristics from the measured mobility \cite{characterisation_high_quality, theory_kkb_model}. Correspondingly, studies of hole mobility can yield valuable information about p-type interface quality, which is instrumental in the pursuit of large numbers of hole spin qubits.
Whereas low temperature mobilities in excess of $30,000\,\text{cm}^2/\text{Vs}$ have been reported for electrons at MOS interfaces \cite{kravchenko1994possible,manufacturing}, reported hole mobilities tend to be an order of magnitude lower \cite{betz2014ambipolar,mueller2015single}. This is somewhat unexpected, given that the often-cited hole transverse effective mass $0.22m_0$ is only slightly larger than the electron transverse effective mass $0.19m_0$ (appendix \ref{appendix:holes}). It is important to identify the cause of this low hole mobility, and whether it reflects a difference in disorder that could be detrimetal to formation of hole qubit devices, or is some other effect. 
\\

\input{figures/IMEC_record}

To study this mobility discrepancy we measure two pairs of comparable n-type and p-type Hall bars. One pair of devices is fabricated on a 300-mm wafer at IMEC, and the other is fabricated using academic cleanroom facilities at UNSW. The IMEC samples show the highest electron and hole mobilities in a thin oxide silicon MOSFET, while the UNSW samples provide a direct comparison between electron and hole mobilities on wafers with nominally identical oxides. Across both pairs of devices, we observe hole mobilities that are roughly an order of magnitude lower than the corresponding electron mobilities. Although the band edge mobilities for electrons and holes are similar, here we show that the mobility difference can be explained by the highly non-parabolic nature of the valence band. This nonparabolicity results in a large enhancement of the hole transverse effective mass at Fermi energies relevant to typical transport and spin qubit measurements \cite{donetti2011}. Using this more accurate treatment of the hole mass, we succeed in explaining the observed differences in mobility and confirm that the oxide quality for MOS devices is comparable for electrons and holes. \\

\section{IMEC electron mobility measurements}

% We measured an n-type device with a $21\,\text{nm}$ oxide and a p-type device with an 8 nm oxide made by IMEC. Both devices are fabricated using the same $30\,\text{mm}$ process \cite{characterisation_high_quality,manufacturing}, on separate wafers.\\

% The Heliox/AttoDRY is awkward to talk about... we actually measure at 1.6K on the Heliox but perform a later 1.6K measurement on the AttoDRY when comparing to the p-type device...

We begin by measuring a high-quality n-type silicon MOSFET with a $21\,\text{nm}$ oxide layer, fabricated by IMEC \cite{characterisation_high_quality,manufacturing}. In all our measurements, we used standard lock-in techniques and a 4-point probe setup, applying a source-drain voltage of $100\,\mu\text{V}$ at $33\,\text{Hz}$. Fig. \ref{fig:record_mobility} illustrates results for the n-type IMEC device. We calculate the carrier density $n$ by measuring the Hall voltage $V_{xy}$ at five magnetic fields from $-0.5\,\text{T}$ to $+0.5\,\text{T}$.\\

The extracted density is shown as a function of gate voltage $V_g$ in Fig. \ref{fig:record_mobility} (a). We find a linear relationship between density and gate voltage, as expected for a MOSFET. Assuming a relative permittivity $\kappa_\text{ox}=3.9$ for the silicon oxide, we calculate an oxide thickness of $t_\text{ox} = 20.8\,\text{nm}$, in agreement with the oxide thickness of $21\,\text{nm}$ specified in the fabrication process.\\

At low carrier densities and low temperatures, there is a deviation in the linear relationship between the gate voltage and the carrier density extracted from the Hall effect. This may be due to the narrow width of the Hall probes, and/or the onset of percolation \cite{percolation}. Similar deviations were observed for all the n-type and p-type devices. Because we still expect the MOS interface to function as a capacitor in this regime, we assume a linear relationship between $n$ and $V_g$ for all further analysis.\\

From the resistivity $\rho_{xx}$ and carrier density, we calculate the carrier mobility $\mu$, shown as a function of density in Fig. \ref{fig:record_mobility} (b). At a temperature of $0.25\,\text{K}$ we find a peak mobility of nearly $40,000\,\text{cm}^2/\text{Vs}$, the highest of any thin oxide ($<100$\,\text{nm}) Si/SiO$_2$ MOSFET. Finally, Fig. \ref{fig:record_mobility} (c) shows the contact resistance and resistivity. The contact resistance is calculated from the difference between the total resistance across the device and the longitudinal resistivity $\rho_{xx}$, appropriately considering the device geometry. At $n=0.4\times10^{12}\,\text{cm}^{-2}$ and $0.25\,\text{K}$, we find a contact resistance of only $3\,\text{k}\Omega$.\\

\section{IMEC electron and hole mobilities}

% First we need to introduce the p-type device again

% By measuring the p-type device from IMEC, we can compare electron and hole mobilities. Because these devices are fabricated in similar processes, we expect the interfaces to be largely comparable, though the different originating wafers and oxide thicknesses may introduce some variability.\\

To compare electron and hole mobilities, we measured a p-type silicon MOSFET from IMEC with a $8\,\text{nm}$ oxide, fabricated in the same process as the $21\,\text{nm}$ n-type MOSFET \cite{manufacturing,characterisation_high_quality}. We expect the interface roughness characteristics to be similar between n-type and p-type devices, as they are fabricated in nominally identical processes with the disorder determined by the average microscopic structure of the interface. The Coulomb impurity scattering could be different between n-type and p-type devices, due to the different oxide thicknesses ($8\,\text{nm}$ and $20\,\text{nm}$), which moves the top surface further from the 2D channel in the n-type devices~\cite{manufacturing}. Because scattering in the high density limit is dominated by interface roughness (section \ref{sec:theory}), we expect the disorder to be comparable within this regime.\\

Figure \ref{fig:IMEC_comparison} shows the mobility of both devices as a function of density, measured at $1.6\,\text{K}$. We find peak electron and hole mobilities of $36,000\,\text{cm}^2/\text{Vs}$ and $2,100\,\text{cm}^2/\text{Vs}$ respectively, the latter being the highest recorded mobility for a p-type silicon MOSFET. Despite this, the hole mobility is still an order-of-magnitude lower than the electron mobility at the same temperature. This is unexpected, given that the transverse component of the hole effective mass $0.22m_0$ is only slightly larger than the electron effective mass $\mt_e = 0.19m_0$.\\

\input{figures/IMEC_comparison}

This discrepancy occurs because the hole effective mass $0.22m_0$ is inapplicable at the Fermi energies occurring in this experiment; nonparabolicity of the valence band means that the value $0.22m_0$ is only valid in an extremely limited region around $n=0$. While such nonparabolicity would generally require a more sophisticated calculation~\cite{marcellina2017spin}, Ref. \cite{donetti2011} shows how multiple density-dependent effective masses may be used to model transport in these conditions. At densities of $n \sim 10^{12}\,\text{cm}^{-2}$, which are relevant to our measurements and hole-spin qubit operation \cite{scott2018six}, a transverse effective mass $\mt_h$ ranging from $0.6m_0$ to $0.7m_0$ and a density-of-states effective mass $\mdos_h$ ranging from $0.7m_0$ to $0.9m_0$ may be introduced (see appendix \ref{appendix:holes} and \ref{appendix:mobility_model}).\\

We investigate if this effective mass can explain the difference between the measured electron and hole mobilities. Mobility is approximately related to effective mass via $\mu \propto 1/m^2$, where mass dependence arises due to (a) the relationship between mobility and momentum relaxation time, and (b) the relationship between momentum relaxation time and scattering matrix elements. Note that this approximation is suited to the high-density limit where temperature dependent screening (which introduces additional $m$ dependence) becomes unimportant.\\

The dashed curve in Fig. \ref{fig:IMEC_comparison} shows the hole mobility that is expected by scaling the electron mobility with the square of the band-edge effective masses, that is $\mu_e \times (m_e^t / 0.22 m_0)^2$. Similarly, the dash dotted curve in Fig. \ref{fig:IMEC_comparison} also shows the expected hole mobility, but using the heavier hole masses arising from the band non-parabolicity. The latter curve is a much closer fit to the measured hole mobilities, suggesting that the low hole mobility is a result of the large density-dependent hole mass.

% The dashed and dash dotted curves in Fig. \ref{fig:IMEC_comparison} show the hole mobilities that are expected by scaling the electron mobility with the square of the effective mass ratio, i.e. $\mu_e \times (m_e/m_h)^2$. The dashed line uses the conventional band-edge mass of $0.22m_0$, and is significantly higher than the measured hole mobility. The dash-dotted line uses the heavier hole mass arising from band non-parabolicity, and is a close fit to the measured hole data. This suggests that the low hole mobility is a result of the large density dependent hole mass.

\section{Theoretical mobility analysis}
\label{sec:theory}

To understand the microscopic factors limiting the mobility we use the theoretical analysis of Kruithof, Klapwijk and Bakker (KKB) \cite{theory_kkb_model, characterisation_high_quality}. The KKB model considers scattering from charged impurities and roughness at the MOS interface, as illustrated in Fig. \ref{fig:kkb}.\\

\begin{figure}
    {\linespread{1}\input{figures/cartoon.tikz}}
    \caption{Schematic diagram of a silicon MOSFET. A sheet of charge carriers flows beneath the Si/SiO$_2$ interface. Charged impurities (characterised by $N_C$) and interface roughness (characterised by $\Delta$ and $L$) cause the charge carriers to scatter. A depletion layer wth density $N_\text{dep}$ forms beneath the system.}
    \label{fig:kkb}
\end{figure}

Five different parameters are used to characterise the MOSFET. First is the density of charged impurities, denoted by $N_C$. For simplicity these charges are assumed to lie exactly at the interface, and dominate the scattering at low carrier densities. Next, the interface roughness is described by $\Delta$ and $L$, which represent the amplitude and correlation length respectively. At higher carrier densities the confining electric field pulls the carriers up against the Si/SiO$_2$ interface, increasing the strength of the interface roughness scattering.\\

\input{figures/Diraq_fitting}

Depletion charge in the silicon substrate is also considered. This charge increases the electric field at the interface, amplifying the effect of interface roughness scattering. The increase in electric field also changes the shape of the confining potential and the Fang-Howard wavefunction, affecting both screening and charged impurity scattering. The two-dimensional depletion charge density is given by $N_\text{dep}$, with the sign of the depletion charge taken to match the mobile charge carriers.\\

Finally, the band-tailing parameter $\sigma$ models the reduction in available states at low carrier densities $n$. Here we leave it as a free parameter, but it might be expected to be related to the disorder at low density, arising from the charged impurity density $N_C$ \cite{theory_kkb_model,arnold_band_tails}.\\

When generalising the model to hole mobilities, the effective mass and absence of valley degeneracy must be considered. The density-dependent effective mass defined in Ref. \cite{donetti2011} is straightforward to incorporate. The larger effective mass acts to slow down the carriers, and increase the number of states they scatter into. The absence of valley degeneracy also reduces mobility, by reducing the effects of screening via the density of states. However, the overall effect on screening is not large, as the increased effective mass and absence of valley degeneracy cancel out somewhat. We provide full details of the generalised model in appendix \ref{appendix:mobility_model}.

\section{UNSW electron and hole mobilities}

To unambiguously prove that the difference in mobility between electrons and holes is due to the density dependent hole mass, we fabricated and tested a second set of n-type and p-type Hall bars. These devices were fabricated at UNSW in a laboratory setting. Importantly, both devices were fabricated with a nominally identical $8\,\text{nm}$ oxide thickness. As such, we expect that their interfaces are described by a similar set of parameters in the KKB model (generalised using the density-dependent hole mass).\\

Mobility curves for the n-type and p-type UNSW devices are shown in Fig. \ref{fig:Diraq_fitting} (a) and (b) respectively. Similar to before, there is roughly an order-of-magnitude difference between the measured n-type and p-type mobilities. We fit the generalised KKB model to both the n-type and p-type curves simultaneously, under the constraint that $N_C$, $\Delta$, $L$ and $\sigma$ are the same for both mobility curves. The fitted curves are shown in Fig. \ref{fig:Diraq_fitting}, and the parameters are summarised in table \ref{table:Diraq_fitting}. We succeed in fitting to the both electron and hole mobility curves using these common parameters, demonstrating that the density-dependent hole mass explains the order-of-magnitude difference in mobility. Note that the UNSW devices have negligible temperature dependence, with less than a $2\%$ variation in peak mobility between $1.6\,\text{K}$ and $5.6\,\text{K}$ measurements. Therefore, we ignore temperature in our theoretical modelling.\\

% We fit the generalised KKB model to both the n-type and p-type curves simultaneously, under the constraint that $N_C$, $\Delta$, $L$ and $\sigma$ are shared between both fits. Using this common set of parameters, the model explains the difference in mobility when the large density-dependent hole mass is used. The fitted theory curves are shown in Fig. \ref{fig:Diraq_fitting}, and the parameters are summarised in table \ref{table:Diraq_fitting}. Note that the UNSW devices have negligible temperature dependence, with less than a $2\%$ variation in peak mobility between $1.6\,\text{K}$ and $5.6\,\text{K}$ measurements. Therefore, we ignore temperature in our theoretical modelling.\\

The only parameter we allowed to vary between devices was the depletion charge density, which is zero for the n-type device and slightly non-zero for the p-type device. This is consistent with slight n-type doping in the silicon substrate. While some depletion charge seems necessary to explain the difference in mobility, zeroing this parameter increases the p-type device mobility by only $20\%$.\\

\section{Magnetoresistance measurements}

\begin{figure*}
    \includegraphics[width=\textwidth]{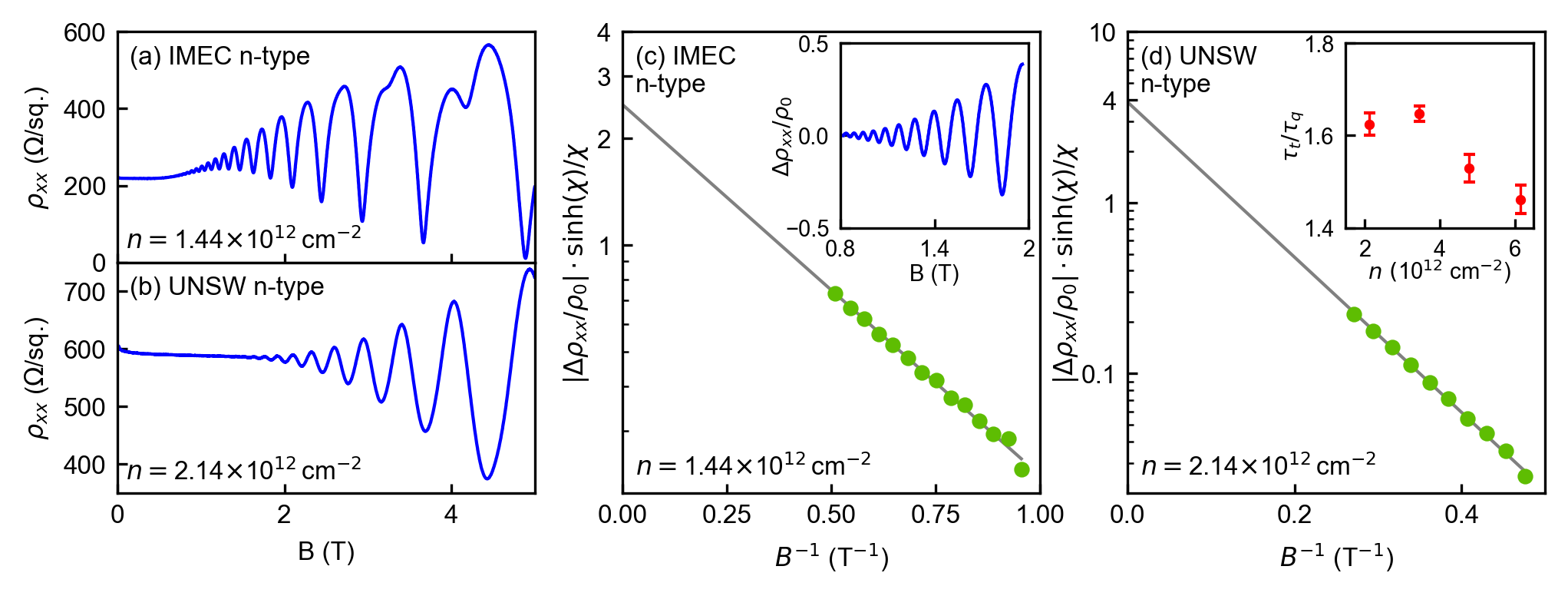}
    \vspace{-2em}
    \caption{(a) Longitudinal resistivity $\rho_{xx}$ as a function of magnetic field $B$ at $1.6\,\text{K}$ and $V_g = 1.5\,\text{V}$ for the IMEC device. The background magnetoresistance shows an unusual upward curvature around $1\,\text{T}$. Same plot at $V_g = 1.0\,\text{V}$ for the UNSW device. Both plots share the same x-axis. (c) Dingle plot of the IMEC device, corresponding to the sweep in (a). The inset shows the scaled resistivity $\Delta \rho_{xx}/\rho_0$ after a background polynomial has been subtracted form the raw $\rho_{xx}$ curve. (d) Dingle plot of the UNSW device, corresponding to the sweep in (b). The inset shows the Dingle ratio $\tau_t/\tau_q$, which decreases as the carrier density increases.}
    \label{fig:shubs}
\end{figure*}
Finally, to build upon the mobility measurements, for completeness we also examine the Shubnikov-de Haas oscillations in the magnetoresistance of the n-type IMEC and UNSW devices for multiple carrier densities, at a temperature of $1.6\,\text{K}$. (We did not observe significant oscillations in our p-type devices at $T=1.6\,\text{K}$ and magnetic fields up to $B=5\,\text{T}$.) These measurements allow us to independently confirm the carrier density, determine the quantum (Dingle) scattering time for the 2D electrons, and examine how the ratio of the quantum and momentum relaxation times varies with carrier density~\cite{dingle2009alex,Wang2013}\\

Figures \ref{fig:shubs} (a) and (b) show these oscillations in the IMEC and UNSW samples. The onset of these oscillations occurs earlier in the IMEC sample than in the UNSW one, which is expected due to the higher mobility. Spin splitting is observed in the IMEC sample at roughly $3\,\text{T}$. An unusual feature of the IMEC device is the positive background magnetoresistance around $1\,\text{T}$. By contrast, both the UNSW device and other silicon MOSFETs \cite{industrial_quantum_transport,characterisation_high_quality,manufacturing} have a flat or negative magnetoresistance.\\

To extract the quantum scattering lifetime from the oscillations, we compare the measured values of $\rho_{xx}$ to the theoretical values predicted by equation \ref{shubs} \cite{coleridge1989}. The resistivity is related to the magnetic field $B$ through the cyclotron frequency $\omega_c$, and the carrier density is related to the Fermi energy via the relations $n = D \cdot \varepsilon_F$ and $D = g_v \mt_e/(\pi \hbar)^2$ where $D$ is the density of states. The quantum lifetime $\tau_q$ represents the amount of time between collisions that dephase the electrons. Note that this equation is valid for two-dimensional systems \footnote{In ref. \cite{industrial_quantum_transport} the 3D variant of the Lifshitz-Kosevich formula, valid for 3D bulk materials, has been used to analyse their 2D electron system.}.

\begin{equation}
\begin{aligned}
&\frac{\Delta \rho_{xx}}{\rho_0} = (-4) \frac{\chi}{\sinh \chi} \cos \left( \frac{2\pi\varepsilon_F}{\hbar \omega_c} \right) \exp \left( -\frac{\pi}{\omega_c \tau_q} \right) \\
&\qquad\qquad \chi = \frac{2\pi^2 k_BT}{\hbar \omega_c}, \quad \omega_c = \frac{eB}{m_e^t}
\end{aligned}
\label{shubs}
\end{equation}

We subtract a background polynomial from the raw $\rho_{xx}$ data to reveal the oscillatory component $\Delta \rho_{xx}$. This process is exemplified in the inset of Fig. \ref{fig:shubs} (c) for the IMEC device. The frequency of the SdH oscillations provides an independent measure of the carrier density. In all our measurements, this agrees with the density obtains via the low-field Hall effect to within 2\%.\\

\input{figures/shubs_table}

We use Dingle plots to calculate the quantum lifetime \cite{coleridge1989,coleridge1991}. These are produced by plotting the magnitude of the peaks and valleys in the temperature-adjusted resistivity $|\Delta \rho_{xx}/\rho_0 | \times \sinh\chi/\chi$ against the inverse field $1/B$. This quantity should appear linear on a log plot, and from its gradient we can extract $\tau_q$. Note that equation \ref{shubs} implies that the intercept of the resultant curve should be $4$ \cite{coleridge1989}. A good Dingle plot should have this intercept, as shown in Fig. \ref{fig:shubs} (d).\\

Good Dingle plots are obtained for the UNSW device at all carrier densities. However, the intercepts for the IMEC devices fall below $4$ for each sweep, as shown in Fig. \ref{fig:shubs} (c). To determine if this significantly affects the extracted quantum lifetimes of the IMEC samples, we consider Dingle plots whose intercepts are constrained to pass through $4$. This produces quantum lifetimes which are $20$--$30\%$ smaller, but still show the same trend with carrier density \footnote{We leave the explanation of the bad IMEC Dingle plots to future work. They may be related to a sub-optimal Hall bar geometry which distorts equipotential lines at finite $B$. A possible cause may also be related to the positive background magnetoresistance} \footnote{See ref. \cite{coleridge1991} for discussion of Dingle plots with bad intercepts. In some circumstances, calculating $\tau_q$ from such a Dingle plot is still valid}.\\

From the quantum lifetime, we calculate the Dingle ratio $\tau_t/\tau_q$ which characterises the typical range of scatterers from the 2DEG. These are summarised in table \ref{table:shubs} for both devices. Larger Dingle ratios indicate that long-range scattering is dominant \cite{dingle2009alex}. For both devices, we find that the Dingle ratio decreases with density. This ratio is plotted for the UNSW device in the inset of Fig. \ref{fig:shubs} (d). This decreasing ratio is expected in light of scattering theory; at higher densities, short-range interface roughness scattering becomes the dominant scattering mechanism.

\section{Conclusion}

We measured electron and hole mobilities from two pairs of similarly fabricated silicon MOSFETs, reporting the highest measured mobilities in a thin oxide device. Across both samples we observe hole mobilities that are roughly an order-of-magnitude lower than those of electrons. We show that this is due to valence band nonparabolicity, which leads to a transverse effective mass much larger than $0.22m_0$ at the Fermi energies present in MOSFETs and qubits. Finally, we measured Shubnikov-de Haas oscillations, which we used to extract Dingle ratios and probe the typical range of scatterers in each device.

\begin{acknowledgments}
This work was supported by the Australian Research Council through grant LP200100019 and the  US Army Research Office through W911NF-23-10092. K.W.C. and A.R.H. acknowledge ARC Fellowships IM230100396 and IL230100072, co-funded by Diraq Pty Ltd, and A.S.D. acknowledges ARC Fellowship FL190100167. UNSW devices were made at the New South Wales node of the Australian National Fabrication Facility. 

%    \item Funded by Australian Research Council, co-funded by Diraq Pty Ltd, IL230100072. {\color{blue} (Alex industry laureate.)}
%    \item Funded by Australian Research Council, LP200100019 {\color{blue} (QED group I assume.)}
%    \item Funded by Australian Research Council, IM230100396 {\color{blue} (Kok Wai Industry Fellowship grant... {\color{red}is this co-funded by Diraq?})}
%    \item Funded by Australian Research Council, FL190100167 {\color{blue} (Dzurak's email)}
%    \item Funded by US Army Research Office, W911NF-23-10092 {\color{blue} (Dzurak's email)}
%    \item UNSW devices were made at the New South Wales node of the Australian National Fabrication Facility.  {\color{blue} (Dzurak's email)}
 %   \item {\color{red} No grants related to IMEC?}
%    \item We would like to thank Feixiang Xiang, Abhay Gupta and Krittika Kumar for valuable assistance in the lab.
%\end{itemize}
\end{acknowledgments}

\section*{Competing interests}
A.S.D. is the CEO and a director of Diraq Pty Ltd. K.W.C., F.E.H., W.H.L. and A.S.D. declare equity interest in Diraq Pty Ltd. The remaining authors declare no competing interests.

\appendix

\section{Device stock numbers}
\label{appendix:stock_numbers}

\begin{table}[H]
\begin{ruledtabular}
    \begin{tabular}{ccccc}
        Manufacturer & Oxide (nm) & Carrier & Identifier & \\
        \hline
        IMEC       & 21 & n-type & \texttt{AL210563\_D06}  \\
        IMEC       & 8  & p-type & \texttt{AL210563\_D07}  \\
        Diraq/UNSW & 8  & n-type & \texttt{6ND03}  \\
        Diraq/UNSW & 8  & p-type & \texttt{6NI01}  \\
    \end{tabular}
\end{ruledtabular}
\caption{Stock numbers and information about the devices. The sign of the charge carriers is set by the doping of the ohmic contacts.}
\label{table:batch}
\end{table}

\section{Effective masses of electrons and holes}
\label{appendix:holes}

The effective mass approximation (EMA) assumes a quadratic relationship between the energy and wavevector of the charge carriers. This approximation simplifies the calculation of transport properties, and provides adequate accuracy in the conduction band of silicon. For the (100) crystal orientation of our devices, two components of the electron effective mass must be considered \cite{stern1967}. First is the transverse component $\mt_e = 0.19m_0$, which largely determines the transport properties of the system. It plays a direct role in determining the density of states, scattering rate and carrier mobility. Second is the vertical component $\mz_e = 0.98m_0$ which determines the energy levels of the quantised subbands that the electrons occupy.\\

In 2D systems, the silicon valence band is split into heavy hole and light hole components. Transport in silicon MOSFETs at low temperatures is dominated by the heavy hole component \cite{donetti2011}. Ref. \cite{winkler_book} (p. 48) provide simple calculation instructions for the transverse and vertical components of the heavy hole effective mass. The values obtained are $\mt_h = 0.22m_0$ and $\mz_h = 0.28m_0$.\\

\input{figures/donetti}

However, because the valence band is significantly nonparabolic, these hole effective masses are only valid in an extremely limited region around the band-edge. In particular, this nonparabolicity is highly relevant for typical Fermi energies occuring in our devices.\\

Ref. \cite{donetti2011} introduce density-dependent effective masses that model the transport of holes. For heavy holes in (100) oriented silicon, they introduce three such effective masses, shown in Fig. \ref{fig:donetti}. The first of these is the \textit{density of states} effective mass $\mdos_h$ which ranges from $0.7m_0$ to $0.9m_0$ at typical carrier densities of $10^{11}\,\text{cm}^{-2}$ to $10^{13}\,\text{cm}^{-2}$. This component is relevant to the density of states and scattering rate calculations. The next is the \textit{transport} effective mass $\mt_h$ which ranges from $0.6m_0$ to $0.7m_0$ over the same densities, and plays a direct role in linking the transport lifetime to mobility. The last is the \textit{quantisation} effective mass $\mz_h = 0.27m_0$ which plays the same role as the vertical effective mass in calculating subband energies. It has negligible dependence on the carrier density.\\

\section{Theoretical mobility model}
\label{appendix:mobility_model}

Here we provide a complete expression of the mobility model used in our calculations. We use the model described in ref. \cite{theory_kkb_model} at $T=0$, generalised to account for hole mobilities.\\

% To begin, we define generalised effective mass parameters for the electrons and holes which take into account the band structure calculations of ref. \cite{donetti2011}. The parameter values are shown in table \ref{table:effective_mass}.\\
Table \ref{table:effective_mass} defined generalised effective masses for the electrons and holes, where the hole masses are the density-dependent expressions from ref. \cite{donetti2011}.\\

\begin{table}[b]
\begin{ruledtabular}
\begin{tabular}{lll}
    Parameter & Electrons ($m_0$) & Holes ($m_0$) \\
    \hline
    $\mt$ & 0.19 & (0.6)--(0.7) \\
    $\mdos$ & \ditto & (0.7)--(0.9) \\
    $\mz$ & 0.98 & 0.28 \\
\end{tabular}
\end{ruledtabular}
\caption{Values of effective mass parameters for both electrons and holes. In this notation, $m^\text{dos}$ and $m^t$ are identical for electrons. The density-dependent hole masses apply for densities between $10^{11}\,\text{cm}^{-2}$ to $10^{13}\,\text{cm}^{-2}$.}
\label{table:effective_mass}
\end{table}

Equation \ref{fermi} introduces the Fermi wave vector $k_F$, Fermi energy $\varepsilon_F$, and density of states $D$. These are computed as a function of the carrier density $n$. The valley degeneracy factor $g_v$ equals $2$ for electrons and $1$ for holes.
\begin{equation}
    k_F = \sqrt{\frac{2\pi n}{g_v}}, \quad D = \frac{g_v \mdos}{\pi \hbar^2}, \quad \varepsilon_F = \frac n D
    \label{fermi}
\end{equation}

We also use the relative electric permittivities of the silicon substrate $\kappa_\text{si}$, the silicon oxide $\kappa_\text{ox}$, and their average $\overline{\kappa}$, as shown in equation \ref{kappa}.
\begin{equation}
    \kappa_\text{si} = 3.9, \quad \kappa_\text{ox} = 11.8, \quad \overline{\kappa} = \frac{\kappa_\text{si} + \kappa_\text{ox}}{2}
    \label{kappa}
\end{equation}

Next we introduce the variational parameter $b$, shown in equation \ref{b}. This parameter is related to the thickness of the carrier wavefunction (which equals $3/b$).
\begin{equation}
    b = \left[ \frac{12 e^2 \mz}{\hbar^2 \epsilon_0 \kappa_\text{si}} \left( N_\text{dep} + \frac{11}{32} n \right) \right]^{1/3}
    \label{b}
\end{equation}

Now we may write down the squared matrix element for charged impurity scattering, which is given by equation \ref{charged_impurity_element}. This equation is the only place where the free parameter $N_C$ enters. The parameter $q$ is the difference in wave vector between the scattered and incoming charge carriers.
\begin{equation}
    |W_C(q)|^2 = N_C \left[ \frac{e^2}{2 \epsilon_0 \overline{\kappa} q} \frac{1}{(1+q/b)^3} \right]^2
    \label{charged_impurity_element}
\end{equation}

Equations \ref{Eavg}, \ref{Ctilde} and \ref{ir} define the interface roughness matrix element. To begin, we introduce the average electric field at the interface $E_\text{avg}$ in equation \ref{Eavg}.
\begin{equation}
    E_\text{avg} = \frac{e}{\epsilon_0 \kappa_\text{si}} \left(N_\text{dep} + \frac 1 2 n \right)
    \label{Eavg}
\end{equation}
Next, equation \ref{Ctilde} defines $\widetilde{C}(q)$, the Fourier transform of the correlation function of the interface roughness. This equation is the only place where the free parameters $\Delta$ and $L$ enter.
\begin{equation}
    \widetilde{C}(q) = \frac{\pi \Delta^2 L^2}{(1 + q^2 L^2/2)^{3/2}}
    \label{Ctilde}
\end{equation}

Finally, we define the squared matrix element for the interface roughness in equation \ref{ir}.
\begin{equation}
    |W_{IR}(q)|^2 = (e E_\text{avg})^2 \widetilde{C}(q)\\[0.7em]
\label{ir}
\end{equation}

Equations \ref{interaction_potentail}, \ref{Gq} and \ref{eps} define the dielectric screening function. We begin by introducing $\widetilde{U}_\text{int}(q)$, which is the Fourier transform of the interaction potential of the charge carriers, and is defined by equation \ref{interaction_potentail} \footnote{The interaction potential in ref. \cite{theory_kkb_model} (eq. 13) uses the silicon permittivity $\kappa_\text{si}$ rather than the average permittivity $\overline{\kappa}$ \cite{stern1967,ando1977screening}. (The usage of $\overline{\kappa}$ correctly accounts for interactions involving image charges in the silicon oxide, as well as charges in the substrate)}.

% Next, we focus on the dielectric screening function $\epsilon(q)$, shown in equation \ref{eps}.
% \begin{equation}
%     \epsilon(q) = 1 + D \cdot \widetilde{U}_\text{int}(q) (1 - G(q)) \text{erf} \left(\frac{\varepsilon_F}{\sigma \sqrt 2} \right)
%     \label{eps}
% \end{equation}

% The first new term here is $\widetilde{U}_\text{int}(q)$. This is the Fourier transform of the interaction potential of the charge carriers \footnote{The interaction potential in ref. \cite{theory_kkb_model} (eq. 13) uses the silicon permittivity $\kappa_\text{si}$ rather than the average permittivity $\overline{\kappa}$ \cite{stern1967,ando1977screening}. (The usage of $\overline{\kappa}$ correctly accounts for interactions involving image charges in the silicon oxide, as well as charges in the substrate)}, and is given by equation \ref{interaction_potentail}.

\begin{equation}
\begin{aligned}
    &\widetilde{U}_\text{int}(q) = \frac{e^2}{2 \epsilon_0 \overline{\kappa} q} F(q), \,\,\,\text{where}\\[1em]
    F(q) = &\,\frac{1}{16}\left(1 + \frac{\kappa_\text{ox}}{\kappa_\text{si}} \right) \frac{(8 + 9q/b + 3 q^2/b^2)}{(1+q/b)^3} \\[0.7em]
    &\!\!\!\!+ \frac 1 2 \left( 1 - \frac{\kappa_\text{ox}}{\kappa_\text{si}} \right) \frac{1}{(1 + q/b)^6}.\\[1em]
\end{aligned}
\label{interaction_potentail}
\end{equation}

The form factor $F(q)$ in equation \ref{interaction_potentail} accounts for the nonzero thickness of the carrier wavefunction. (Note that $F(q) \rightarrow 1$ as $1/b \rightarrow 0$.)\\

% The second new term is the factor $G(q)$ shown in equation \ref{Gq}, which is a correction to the screening at low carrier densities.

Next, equation \ref{Gq} defined the factor $G(q)$, which is a correction to the screening at low carrier densities.

\begin{equation}
    G(q) = \frac{1}{2 g_v} \frac{q}{\sqrt{q^2 + k_F^2}}
    \label{Gq}
\end{equation}

The dielectric screening function is then defined in \ref{eps}. The term $\text{erf}(\varepsilon_F/\sigma \sqrt{2})$ accounts for the reduction in density-of-states at low carrier densities. (The error function is an approximation to a more detailed self-consistent calculation given in ref. \cite{theory_kkb_model}.)\\

\begin{equation}
    \epsilon(q) = 1 + D \cdot \widetilde{U}_\text{int}(q) (1 - G(q)) \text{erf} \left(\frac{\varepsilon_F}{\sigma \sqrt 2} \right)
    \label{eps}
\end{equation}

The scattering matrix elements and dielectric screening function are combined to obtain the $T=0$ scattering rate, shown in equation \ref{appendix:transport_lifetime}. We integrate over the scattering angle $\theta$ in this expression, which we found suitable for our numerical methods.

\begin{equation}
\label{appendix:transport_lifetime}
\begin{aligned}
\frac{1}{\tau_t} &= \frac{\mdos}{\pi \hbar^3} \!\int_0^\pi \!\! d\theta \, (1 - \cos \theta) \frac{|W_C(q)|^2 + |W_{IR}(q)|^2}{\epsilon(q)^2}\\[0.7em]
&\quad \qquad \qquad q = k_F \sqrt{2(1 - \cos\theta)}
\end{aligned}
\end{equation}

% To take temperature into account, we use the first order approximation shown in equation \ref{temp}.

% \begin{widetext}
% \begin{equation}
%     \frac{1}{\tau_t(T)} = \frac{1}{\tau_t(0)} +  (\ln 2) k_BT  k_F^2 \frac{|W_C(q_\text{max})|^2 + |W_{IR}(q_\text{max})^2 |}{\hbar \varepsilon_F^2 } \frac{\epsilon(q_\text{max}) - 1}{\epsilon(q_\text{max})^3}, \quad q_\text{max} = 2k_F
% \label{temp}
% \end{equation}
% \end{widetext}

Finally, the carrier mobility $\mu$ is related to the transport lifetime via equation \ref{mu}.

\begin{equation}
    \mu = \frac{e \tau_t}{\mt}.
    \label{mu}
\end{equation}

\clearpage
\bibliography{references}

\end{document}

%% file: figures/IMEC_record.tex
\begin{figure*}
\includegraphics[width=\textwidth]{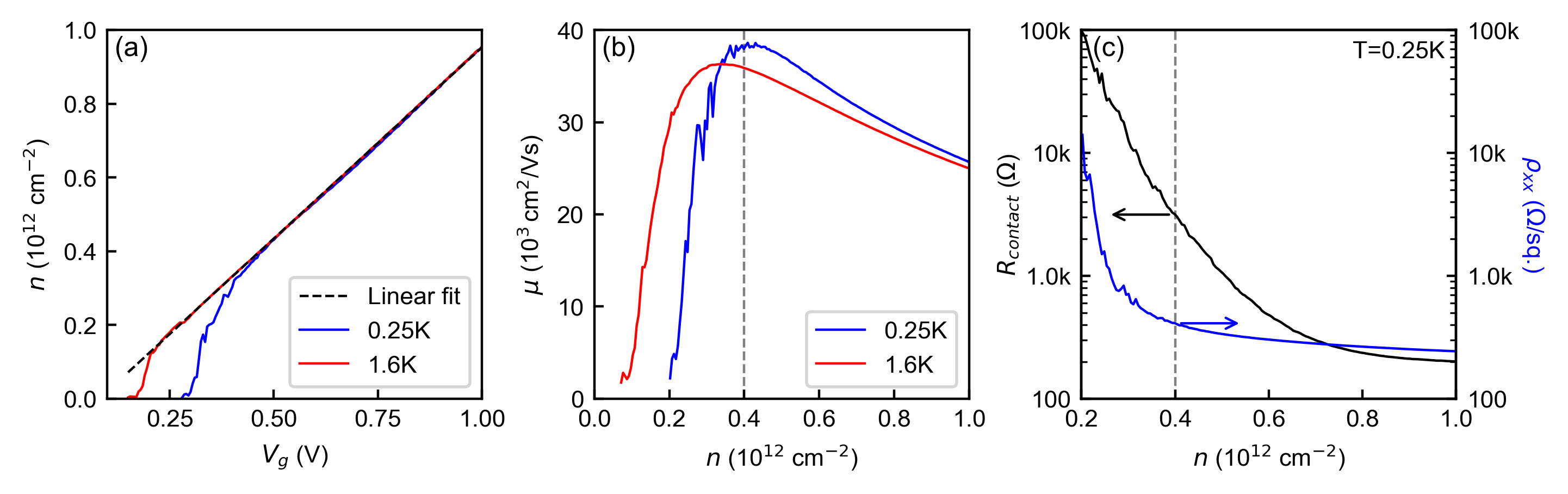}
\vspace{-2em}
\caption{(a) Carrier density $n$ as a function of gate voltage $V_g$ at temperatures $0.25\,\text{K}$ and $1.6\,\text{K}$. A linear fit to the $1.6\,\text{K}$ data yielding an oxide thickness of $t_\text{ox} = 20.8 \,\text{nm}$ is included. (Both curves produce virtually identical fits.) (b) Carrier mobility $\mu$ as a function of density at $0.25\,\text{K}$ and $1.6\,\text{K}$. A peak mobility $\mu = 38,600\,\text{cm}^2/\text{Vs}$ at $0.25\,\text{K}$ is achieved at a density $n\simeq0.4\times10^{12}\,\text{cm}^{-2}$. (c) Contact resistance $R_\text{contact}$ and resistivity $\rho_{xx}$ measured as a function of density at $0.25\,\text{K}$. At the peak mobility, $R_\text{contact} = 3.2\,\text{k}\Omega$ and $\rho_{xx} = 410\,\Omega/\text{sq.}$.}
\label{fig:record_mobility}
\end{figure*}

%% file: figures/IMEC_comparison.tex
\begin{figure}
    % \hspace{-0.5em}
    \includegraphics[width=0.45\textwidth]{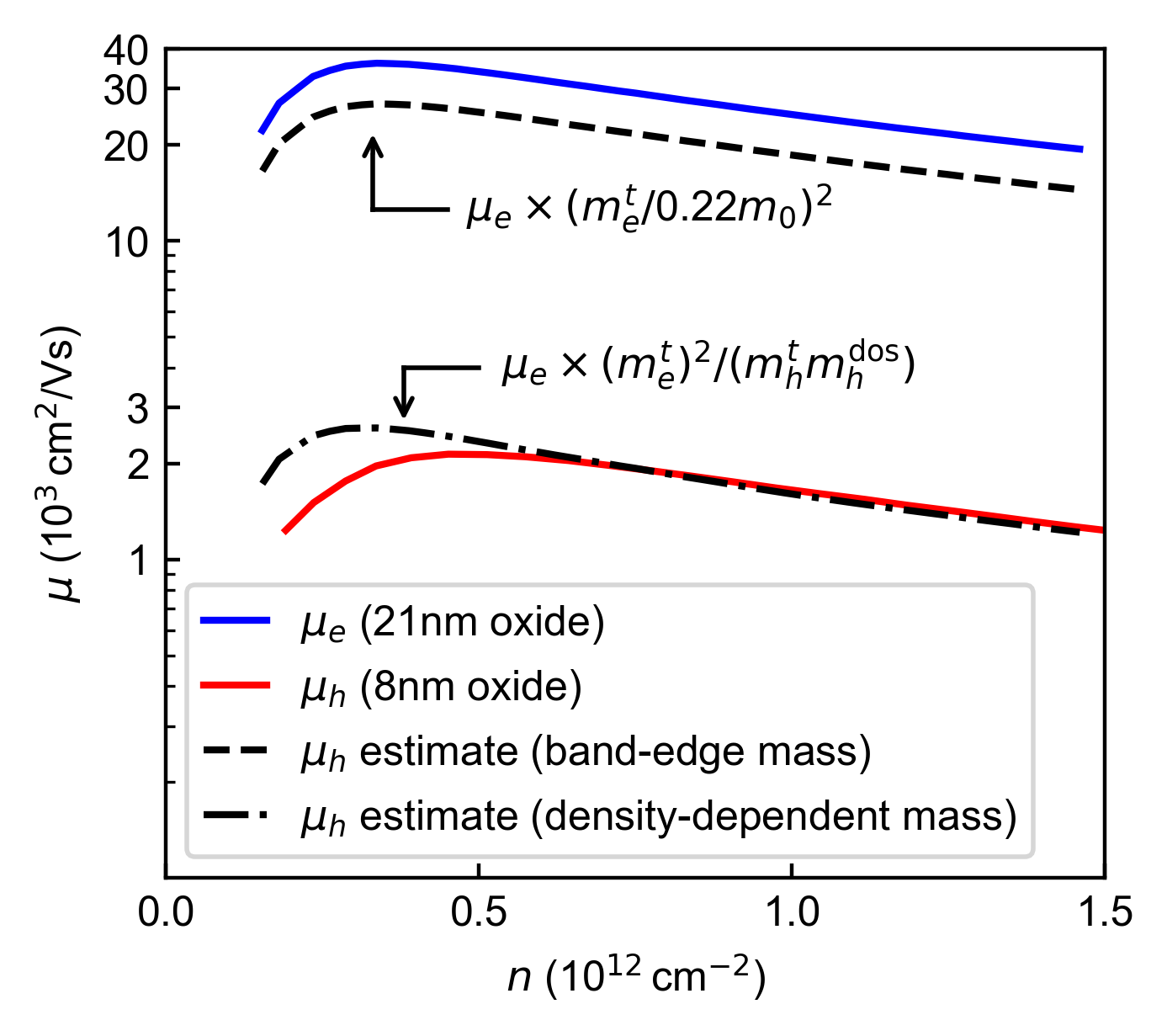}
    \vspace{-1em}
    \caption{Measured electron mobility $\mu_e$ and hole mobility $\mu_h$ of the IMEC devices at $1.6\,\text{K}$, with peak mobilities of $36,000\,\text{cm}^2/\text{Vs}$ and $2,100\,\text{cm}^2/\text{Vs}$ respectively. The dashed and dash-dotted lines show the predicted mobility for holes using the band-edge effective mass $0.22m_0$ and the density-dependent effective masses from Ref. \cite{donetti2011} respectively.}
    \label{fig:IMEC_comparison}
\end{figure}

%% file: figures/cartoon.tikz
\tikzset{every picture/.style={line width=0.75pt}} %set default line width to 0.75pt        

\begin{tikzpicture}[x=0.75pt,y=0.75pt,yscale=-1,xscale=1]

%Shape: Rectangle [id:dp6628100064016991] 
\draw  [draw opacity=0][fill={rgb, 255:red, 214; green, 214; blue, 214 }  ,fill opacity=1 ] (214.46,156.1) -- (329.73,156.1) -- (329.73,211.13) -- (214.46,211.13) -- cycle ;
%Shape: Rectangle [id:dp3479509816108213] 
\draw  [draw opacity=0][fill={rgb, 255:red, 54; green, 64; blue, 248 }  ,fill opacity=1 ] (214.96,155.86) -- (329.23,155.86) -- (329.23,161.73) -- (214.96,161.73) -- cycle ;
%Shape: Rectangle [id:dp23120070390076752] 
\draw  [color={rgb, 255:red, 0; green, 0; blue, 0 }  ,draw opacity=1 ][fill={rgb, 255:red, 248; green, 197; blue, 124 }  ,fill opacity=1 ] (214.46,116.1) -- (329.73,116.1) -- (329.73,132.1) -- (214.46,132.1) -- cycle ;
%Straight Lines [id:da9550994296197788] 
\draw [color={rgb, 255:red, 54; green, 64; blue, 248 }  ,draw opacity=1 ]   (198.48,182.57) -- (219.48,161.57) ;
%Straight Lines [id:da6167989201708539] 
\draw [color={rgb, 255:red, 54; green, 64; blue, 248 }  ,draw opacity=1 ]   (205.48,182.57) -- (226.48,161.57) ;
%Straight Lines [id:da9652753614274192] 
\draw [color={rgb, 255:red, 54; green, 64; blue, 248 }  ,draw opacity=1 ]   (212.48,182.57) -- (233.48,161.57) ;
%Straight Lines [id:da869363511192998] 
\draw [color={rgb, 255:red, 54; green, 64; blue, 248 }  ,draw opacity=1 ]   (219.48,182.57) -- (240.48,161.57) ;
%Straight Lines [id:da4711990734545426] 
\draw [color={rgb, 255:red, 54; green, 64; blue, 248 }  ,draw opacity=1 ]   (226.48,182.57) -- (247.48,161.57) ;
%Straight Lines [id:da40986642928926176] 
\draw [color={rgb, 255:red, 54; green, 64; blue, 248 }  ,draw opacity=1 ]   (233.48,182.57) -- (254.48,161.57) ;
%Straight Lines [id:da2538818496758468] 
\draw [color={rgb, 255:red, 54; green, 64; blue, 248 }  ,draw opacity=1 ]   (240.48,182.57) -- (261.48,161.57) ;
%Straight Lines [id:da0092783171919808] 
\draw [color={rgb, 255:red, 54; green, 64; blue, 248 }  ,draw opacity=1 ]   (247.48,182.57) -- (268.48,161.57) ;
%Straight Lines [id:da7086901627242264] 
\draw [color={rgb, 255:red, 54; green, 64; blue, 248 }  ,draw opacity=1 ]   (254.48,182.57) -- (275.48,161.57) ;
%Straight Lines [id:da340067729358124] 
\draw [color={rgb, 255:red, 54; green, 64; blue, 248 }  ,draw opacity=1 ]   (261.48,182.57) -- (282.48,161.57) ;
%Straight Lines [id:da10929572284049249] 
\draw [color={rgb, 255:red, 54; green, 64; blue, 248 }  ,draw opacity=1 ]   (268.48,182.57) -- (289.48,161.57) ;
%Straight Lines [id:da8642346345772964] 
\draw [color={rgb, 255:red, 54; green, 64; blue, 248 }  ,draw opacity=1 ]   (275.48,182.57) -- (296.48,161.57) ;
%Straight Lines [id:da7973308613749563] 
\draw [color={rgb, 255:red, 54; green, 64; blue, 248 }  ,draw opacity=1 ]   (282.48,182.57) -- (303.48,161.57) ;
%Straight Lines [id:da22362532099226762] 
\draw [color={rgb, 255:red, 54; green, 64; blue, 248 }  ,draw opacity=1 ]   (289.48,182.57) -- (310.48,161.57) ;
%Straight Lines [id:da8048817194915234] 
\draw [color={rgb, 255:red, 54; green, 64; blue, 248 }  ,draw opacity=1 ]   (296.48,182.57) -- (317.48,161.57) ;
%Straight Lines [id:da6666033703764298] 
\draw [color={rgb, 255:red, 54; green, 64; blue, 248 }  ,draw opacity=1 ]   (303.48,182.57) -- (324.48,161.57) ;
%Straight Lines [id:da8776562828645191] 
\draw [color={rgb, 255:red, 54; green, 64; blue, 248 }  ,draw opacity=1 ]   (310.48,182.57) -- (331.48,161.57) ;
%Straight Lines [id:da9440962950338869] 
\draw [color={rgb, 255:red, 54; green, 64; blue, 248 }  ,draw opacity=1 ]   (317.48,182.57) -- (338.48,161.57) ;
%Straight Lines [id:da4017378649552663] 
\draw [color={rgb, 255:red, 54; green, 64; blue, 248 }  ,draw opacity=1 ]   (324.48,182.57) -- (345.48,161.57) ;
%Straight Lines [id:da3875530904022628] 
\draw [color={rgb, 255:red, 54; green, 64; blue, 248 }  ,draw opacity=1 ]   (331.48,182.57) -- (352.48,161.57) ;

%Shape: Rectangle [id:dp5453159896967967] 
\draw  [draw opacity=0][fill={rgb, 255:red, 255; green, 255; blue, 255 }  ,fill opacity=1 ] (191.48,159.63) -- (214.96,159.63) -- (214.96,194.57) -- (191.48,194.57) -- cycle ;
%Shape: Rectangle [id:dp5620610797753771] 
\draw  [draw opacity=0][fill={rgb, 255:red, 255; green, 255; blue, 255 }  ,fill opacity=1 ] (329.61,154.6) -- (353.98,154.6) -- (353.98,187.57) -- (329.61,187.57) -- cycle ;
%Shape: Rectangle [id:dp2502099594406144] 
\draw  [draw opacity=0][fill={rgb, 255:red, 214; green, 214; blue, 214 }  ,fill opacity=1 ] (214.81,180.2) -- (329.61,180.2) -- (329.61,186.9) -- (214.81,186.9) -- cycle ;

%Straight Lines [id:da4904042023158963] 
\draw    (341.15,188.52) -- (308.32,171.93) ;
%Shape: Polygon [id:ds41475872599483776] 
\draw  [draw opacity=0][fill={rgb, 255:red, 174; green, 233; blue, 117 }  ,fill opacity=1 ] (293.84,89.72) -- (311.32,82.64) -- (329.25,90.55) -- (345.61,85.8) -- (367.24,87.91) -- (367.44,82.07) -- (366.12,74.43) -- (363.48,67.57) -- (358.99,60.97) -- (353.19,55.7) -- (346.6,51.21) -- (339.21,48.58) -- (331.56,47.52) -- (323.65,48.31) -- (316,50.69) -- (309.14,54.38) -- (303.08,59.92) -- (298.06,66.78) -- (295.16,73.63) -- (293.84,81.55) -- cycle ;
%Shape: Polygon [id:ds7338487068273603] 
\draw  [draw opacity=0][fill={rgb, 255:red, 214; green, 214; blue, 214 }  ,fill opacity=1 ] (293.84,89.97) -- (311.32,82.89) -- (329.45,90.48) -- (345.61,86.05) -- (367.24,88.16) -- (365.33,97.45) -- (362.55,103.94) -- (357.92,110.11) -- (352.05,115.06) -- (345.88,118.45) -- (339.39,120.92) -- (331.67,121.85) -- (325.8,121.54) -- (317.77,119.69) -- (310.98,115.98) -- (306.03,112.58) -- (300.78,107.02) -- (296.46,99.61) -- cycle ;
%Shape: Ellipse [id:dp2598157615630259] 
\draw  [dash pattern={on 3pt off 1.5pt}] (293.99,84.9) .. controls (293.99,64.49) and (310.45,47.95) .. (330.76,47.95) .. controls (351.07,47.95) and (367.54,64.49) .. (367.54,84.9) .. controls (367.54,105.31) and (351.07,121.85) .. (330.76,121.85) .. controls (310.45,121.85) and (293.99,105.31) .. (293.99,84.9) -- cycle ;
%Straight Lines [id:da6940023046710542] 
\draw    (371.15,51.68) -- (306.48,71.75) ;
%Shape: Circle [id:dp006923907867203427] 
\draw  [fill={rgb, 255:red, 255; green, 0; blue, 0 }  ,fill opacity=1 ] (298.84,78.53) .. controls (298.84,75.81) and (301.05,73.59) .. (303.78,73.59) .. controls (306.51,73.59) and (308.72,75.81) .. (308.72,78.53) .. controls (308.72,81.26) and (306.51,83.48) .. (303.78,83.48) .. controls (301.05,83.48) and (298.84,81.26) .. (298.84,78.53) -- cycle ;
%Straight Lines [id:da05938085883858135] 
\draw    (301.31,78.53) -- (306.25,78.53) ;
%Straight Lines [id:da20280304416649808] 
\draw    (303.78,76.06) -- (303.78,81.01) ;
%Shape: Ellipse [id:dp6365844747895035] 
\draw  [fill={rgb, 255:red, 62; green, 119; blue, 255 }  ,fill opacity=1 ] (326.08,82.6) .. controls (326.08,79.87) and (328.3,77.66) .. (331.03,77.66) .. controls (333.76,77.66) and (335.97,79.87) .. (335.97,82.6) .. controls (335.97,85.33) and (333.76,87.54) .. (331.03,87.54) .. controls (328.3,87.54) and (326.08,85.33) .. (326.08,82.6) -- cycle ;
%Straight Lines [id:da6898256696403537] 
\draw    (328.56,82.6) -- (333.5,82.6) ;
%Straight Lines [id:da8861454654056438] 
\draw    (372.15,53.68) -- (331.19,74.83) ;
%Shape: Ellipse [id:dp7301297079173523] 
\draw  [fill={rgb, 255:red, 255; green, 0; blue, 0 }  ,fill opacity=1 ] (352.1,80.43) .. controls (352.1,77.7) and (354.32,75.49) .. (357.04,75.49) .. controls (359.77,75.49) and (361.99,77.7) .. (361.99,80.43) .. controls (361.99,83.16) and (359.77,85.38) .. (357.04,85.38) .. controls (354.32,85.38) and (352.1,83.16) .. (352.1,80.43) -- cycle ;
%Straight Lines [id:da724817977167147] 
\draw    (354.57,80.43) -- (359.52,80.43) ;
%Straight Lines [id:da8789024220819945] 
\draw    (357.04,77.96) -- (357.04,82.9) ;
%Straight Lines [id:da546990493992163] 
\draw    (294.01,90.47) -- (311.11,82.62) -- (330.77,90.62) -- (345.9,86.25) -- (367.26,87.97) ;

%Shape: Polygon [id:ds30175853449499623] 
\draw  [draw opacity=0][fill={rgb, 255:red, 174; green, 233; blue, 117 }  ,fill opacity=1 ] (368.73,129.05) -- (386.21,121.97) -- (404.15,129.88) -- (420.5,125.13) -- (442.13,127.24) -- (442.33,121.4) -- (441.01,113.75) -- (438.37,106.9) -- (433.89,100.3) -- (428.08,95.03) -- (421.49,90.54) -- (414.1,87.91) -- (406.45,86.85) -- (398.54,87.64) -- (390.89,90.02) -- (384.03,93.71) -- (377.97,99.25) -- (372.96,106.11) -- (370.05,112.96) -- (368.73,120.88) -- cycle ;
%Shape: Polygon [id:ds6204576293134996] 
\draw  [draw opacity=0][fill={rgb, 255:red, 214; green, 214; blue, 214 }  ,fill opacity=1 ] (368.73,129.3) -- (386.21,122.22) -- (404.34,129.81) -- (420.5,125.38) -- (442.13,127.49) -- (440.23,136.78) -- (437.45,143.27) -- (432.81,149.44) -- (426.94,154.38) -- (420.77,157.78) -- (414.28,160.25) -- (406.56,161.18) -- (400.69,160.87) -- (392.66,159.02) -- (385.87,155.31) -- (380.93,151.91) -- (375.67,146.35) -- (371.35,138.94) -- cycle ;
%Shape: Ellipse [id:dp9724123680028784] 
\draw  [dash pattern={on 3pt off 1.5pt}] (368.88,124.23) .. controls (368.88,103.82) and (385.34,87.27) .. (405.65,87.27) .. controls (425.96,87.27) and (442.43,103.82) .. (442.43,124.23) .. controls (442.43,144.64) and (425.96,161.18) .. (405.65,161.18) .. controls (385.34,161.18) and (368.88,144.64) .. (368.88,124.23) -- cycle ;
%Straight Lines [id:da9522381526832786] 
\draw    (448.4,129.75) -- (448.4,123.22) -- (448.4,121.68) ;
\draw [shift={(448.4,118.68)}, rotate = 90] [fill={rgb, 255:red, 0; green, 0; blue, 0 }  ][line width=0.08]  [draw opacity=0] (5.36,-2.57) -- (0,0) -- (5.36,2.57) -- (3.56,0) -- cycle    ;
\draw [shift={(448.4,132.75)}, rotate = 270] [fill={rgb, 255:red, 0; green, 0; blue, 0 }  ][line width=0.08]  [draw opacity=0] (5.36,-2.57) -- (0,0) -- (5.36,2.57) -- (3.56,0) -- cycle    ;
%Straight Lines [id:da9298227900821776] 
\draw    (415.29,134.37) -- (391.64,134.37) ;
\draw [shift={(388.64,134.37)}, rotate = 360] [fill={rgb, 255:red, 0; green, 0; blue, 0 }  ][line width=0.08]  [draw opacity=0] (5.36,-2.57) -- (0,0) -- (5.36,2.57) -- (3.56,0) -- cycle    ;
\draw [shift={(418.29,134.37)}, rotate = 180] [fill={rgb, 255:red, 0; green, 0; blue, 0 }  ][line width=0.08]  [draw opacity=0] (5.36,-2.57) -- (0,0) -- (5.36,2.57) -- (3.56,0) -- cycle    ;
%Straight Lines [id:da24393095250605135] 
\draw    (368.91,129.8) -- (386,121.95) -- (405.67,129.95) -- (420.79,125.57) -- (442.16,127.3) ;

%Straight Lines [id:da42283720680810477] 
\draw    (239.15,120.52) -- (222.58,98.35) ;
%Straight Lines [id:da8049486365849536] 
\draw    (329.73,204.1) -- (329.73,156.1) -- (214.46,156.1) -- (214.46,204.1) ;
%Shape: Rectangle [id:dp9960917536154321] 
\draw  [color={rgb, 255:red, 0; green, 0; blue, 0 }  ,draw opacity=1 ][fill={rgb, 255:red, 174; green, 233; blue, 117 }  ,fill opacity=1 ] (214.46,132.1) -- (329.73,132.1) -- (329.73,156.1) -- (214.46,156.1) -- cycle ;
%Straight Lines [id:da09728173606388268] 
\draw  [dash pattern={on 3pt off 1.5pt}]  (304.6,153.82) -- (370.6,133.82) ;
%Shape: Ellipse [id:dp7013479751150151] 
\draw  [dash pattern={on 3pt off 1.5pt}] (286.14,156.22) .. controls (286.14,150.78) and (290.46,146.37) .. (295.79,146.37) .. controls (301.12,146.37) and (305.44,150.78) .. (305.44,156.22) .. controls (305.44,161.66) and (301.12,166.06) .. (295.79,166.06) .. controls (290.46,166.06) and (286.14,161.66) .. (286.14,156.22) -- cycle ;
%Straight Lines [id:da9457837871347932] 
\draw  [dash pattern={on 3pt off 1.5pt}]  (300.1,147.82) -- (314.6,117.82) ;
%Straight Lines [id:da30759808704309033] 
\draw  [dash pattern={on 2.25pt off 1.5pt}]  (214.46,212.13) -- (214.46,204.1) ;
%Straight Lines [id:da23817682324310208] 
\draw  [dash pattern={on 2.25pt off 1.5pt}]  (329.73,212.13) -- (329.73,204.1) ;
%Straight Lines [id:da2925191232927] 
\draw    (221.93,158.63) -- (189.43,177.63) ;

% Text Node
\draw (346.33,182) node [anchor=north west][inner sep=0.75pt]   [align=left] {Depletion region\\($\displaystyle N_{\text{dep}}$)};
% Text Node
\draw (373.73,42.74) node [anchor=north west][inner sep=0.75pt]  [font=\small]  {$N_{C}$};
% Text Node
\draw (216.46,138.1) node [anchor=north west][inner sep=0.75pt]   [align=left] {SiO$\displaystyle _{2}$};
% Text Node
\draw (216.31,186.2) node [anchor=north west][inner sep=0.75pt]   [align=left] {Si};
% Text Node
\draw (180.31,80.7) node [anchor=north west][inner sep=0.75pt]   [align=left] {Top gate};
% Text Node
\draw (452.9,120.08) node [anchor=north west][inner sep=0.75pt]  [font=\small]  {$\Delta $};
% Text Node
\draw (399.19,136.92) node [anchor=north west][inner sep=0.75pt]  [font=\small]  {$L$};
% Text Node
\draw (153.83,184) node [anchor=north west][inner sep=0.75pt]   [align=left] {Charge\\carriers};

\end{tikzpicture}

%% file: figures/Diraq_fitting.tex
\begin{figure*}[ht]
    \centering
    \begin{minipage}[t]{0.64\textwidth}
        \centering
        \includegraphics[width=\textwidth]{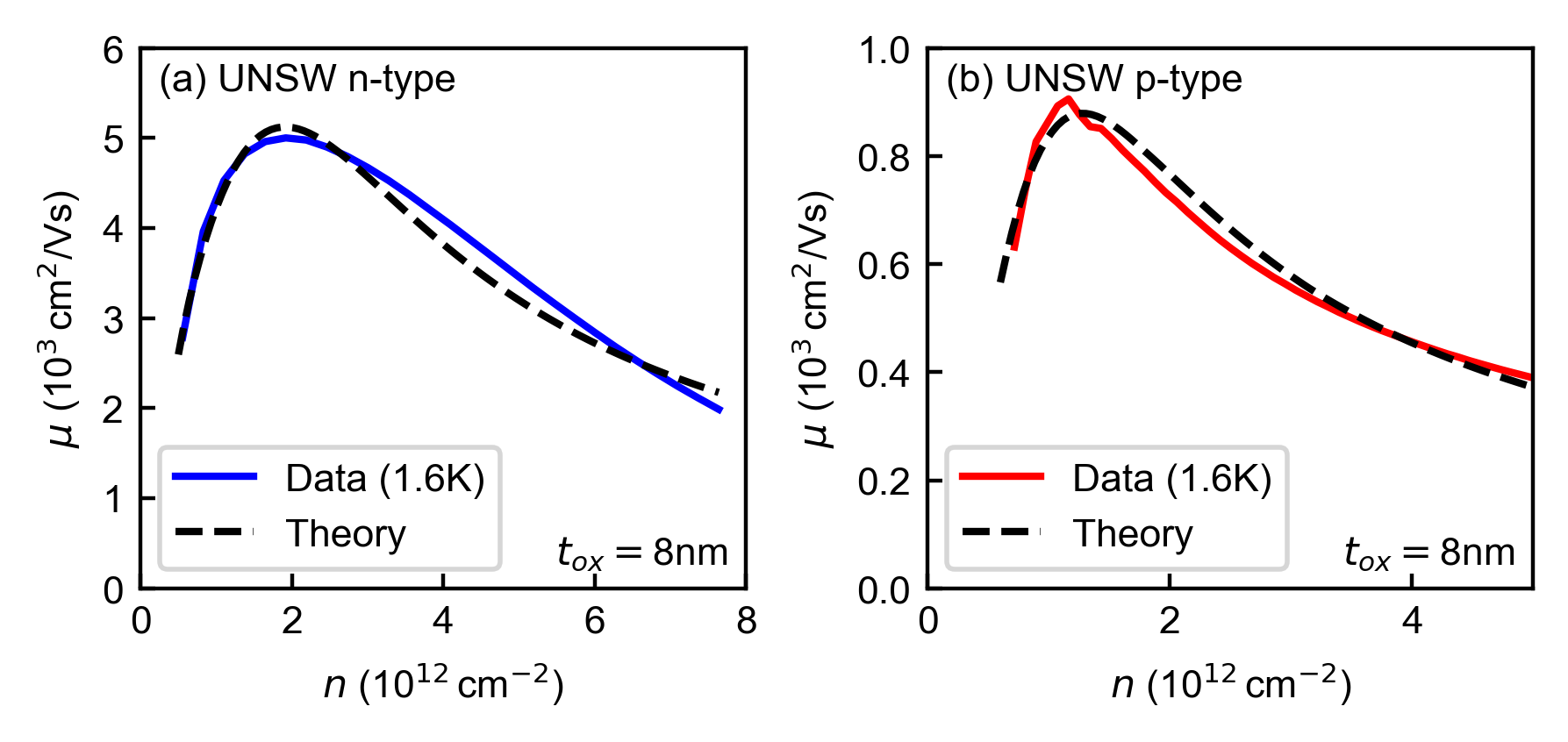}
        \caption{(a) Measured and theoretical carrier mobilities for the UNSW fabricated n-type device. The $T=0$ theoretical model has been fit to both the n-type and p-type mobility curves simultaneously, with the density-dependent hole masses used for the p-type device. (b) Measured and theoretical carrier mobilities for the p-type device.}
        \label{fig:Diraq_fitting}
    \end{minipage}
    \hfill
    \begin{minipage}[t]{0.33\textwidth}
        \centering
        \vspace{-15em}
        \begin{ruledtabular}
            \begin{tabular}{lcc}
            Parameter & n-type & p-type \\
            \hline
            $N_C\,(10^{10}\,\text{cm}^{-2})$            & $5.1 \pm 0.3$   & ---\texttt{"}--- \\
            $\Delta$ (nm)                               & $0.39 \pm 0.01$ & ---\texttt{"}--- \\
            $L$ (nm)                                    & $3.7 \pm 0.4$   & ---\texttt{"}--- \\
            $\sigma$ (meV)                              & $1.7 \pm 0.4$   & ---\texttt{"}--- \\
            $N_\text{dep}\,(10^{12}\,\text{cm}^{-2})$   & $0.00 \pm 0.01$   & $0.15 \pm 0.02$ \\
            \end{tabular}
        \end{ruledtabular}
        \vspace{5.6em}
        \captionof{table}{Parameter values for the fits in figure \ref{fig:Diraq_fitting}. The values of $N_C$, $\Delta$, $L$ and $\sigma$ are identical for both mobility curves.}
        \label{table:Diraq_fitting}
    \end{minipage}
\end{figure*}

%% file: figures/shubs_table.tex
% \begin{table}

% \begin{ruledtabular}
%     \begin{tabular}{ccccc}

%         Device & $V_g$ (V) & $n\,(10^{12}\,\text{cm}^{-2})$ & $\tau_t/\tau_q$ & $\tau_t/\tau_q$ (constr.)\\

%         \hline
%         UNSW             & 1.0 & $2.14(2)$ & 1.62(2) & --- \\
%         ---\texttt{"}--- & 1.5 & $3.49(3)$ & 1.65(2) & --- \\
%         ---\texttt{"}--- & 2.0 & $4.83(5)$ & 1.53(3) & --- \\
%         ---\texttt{"}--- & 2.5 & $6.18(5)$ & 1.46(3) & --- \\
%         \hline
%         IMEC             & 0.7 & $0.63(1)$ & 2.7 & 3.3 \\
%         ---\texttt{"}--- & 1.1 & $1.04(2)$ & 1.9 & 2.4 \\
%         ---\texttt{"}--- & 1.5 & $1.44(3)$ & 1.5 & 1.8 \\
        
%     \end{tabular}
% \end{ruledtabular}

% \caption{Data extracted from the Shubnikov-de Haas oscillations for the UNSW and IMEC n-type devices. The carrier density $n$ is calculated by averaging the values extracted from the Hall and Shubnikov-de Haas oscillations; the uncertainty reflects the difference between these estimates. For the IMEC device, we consider Dingle ratios with an intercept constrained to pass through 4. (This difference dominates all other uncertainties on the Dingle ratio.)}

% \label{table:shubs}
% \end{table}

\begin{table}

\begin{ruledtabular}
    \begin{tabular}{ccccc}

        Device & $V_g$ (V) & $n\,(10^{12}\,\text{cm}^{-2})$ & $\tau_t/\tau_q$ & $\tau_t/\tau_q$ (constr.)\\

        \hline
        UNSW             & 1.0 & $2.14\pm0.02$ & $1.62\pm0.02$ & --- \\
        ---\texttt{"}--- & 1.5 & $3.49\pm0.03$ & $1.65\pm0.02$ & --- \\
        ---\texttt{"}--- & 2.0 & $4.83\pm0.05$ & $1.53\pm0.03$ & --- \\
        ---\texttt{"}--- & 2.5 & $6.18\pm0.05$ & $1.46\pm0.03$ & --- \\
        \hline
        IMEC             & 0.7 & $0.63\pm0.01$ & 2.7 & 3.3 \\
        ---\texttt{"}--- & 1.1 & $1.04\pm0.02$ & 1.9 & 2.4 \\
        ---\texttt{"}--- & 1.5 & $1.44\pm0.03$ & 1.5 & 1.8 \\
        
    \end{tabular}
\end{ruledtabular}

\caption{Data extracted from the SdH oscillations for the UNSW and IMEC n-type devices. The carrier density $n$ is calculated by averaging the values extracted from the Hall effect and SdH oscillations; the uncertainty reflects the difference between these estimates. For the IMEC device, we calculate Dingle ratios where the intercept is constrained to pass through 4. (The difference between these estimates dominates all other sources of uncertainty.)}

\label{table:shubs}
\end{table}

%% file: figures/donetti.tex
\begin{figure}
    \includegraphics[width=0.4\textwidth]{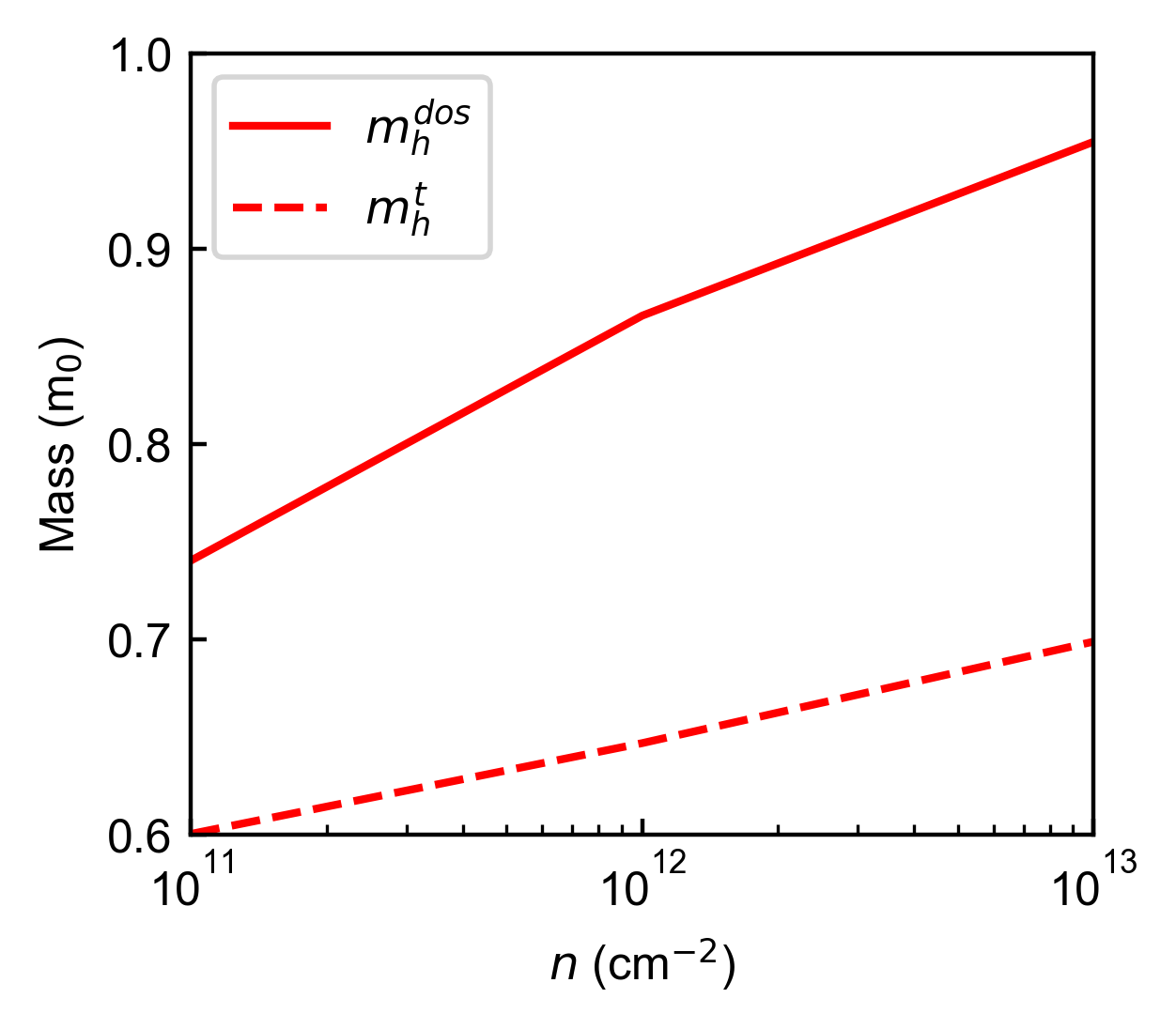}
    \vspace{-1em}
    \caption{Effective mass components for holes as a function of carrier density. The vertical component is given by $\mz_h = 0.28 m_0$. Reproduced from ref. \cite{donetti2011}.}
    \label{fig:donetti}
\end{figure}